\documentclass[twocolumn,amsmath,amssymb,nofootinbib,eqsecnum,tighten]{revtex4}

\usepackage{graphicx}
\usepackage{dcolumn} 
\usepackage{bm}      

\begin{document}


\title{Topological Origin of Zero-Energy Edge States in 
Particle-Hole Symmetric Systems}

\author{
Shinsei RYU$^{1}$ and Yasuhiro HATSUGAI$^{1,2}$
}
\affiliation{
$^1$Department of Applied Physics, University of Tokyo, 7-3-1 Hongo Bunkyo-ku, 
Tokyo 113-8656, Japan \\
$^2$PRESTO, JST, Saitama 332-0012, Japan}


\date{\today}

\begin{abstract}
A criterion to determine
the existence of zero-energy edge states
is discussed 
for a class of particle-hole symmetric Hamiltonians.
A ``loop'' in a parameter space is assigned
for each one-dimensional bulk Hamiltonian,
and its topological properties,
combined with the chiral symmetry,
play an essential role.
It provides
a unified framework
to discuss zero-energy edge modes for several systems
such as fully gapped superconductors,
two-dimensional $d$-wave superconductors,
and graphite ribbons.
A variants of the Peierls instability caused by the presence of edges
is also discussed.
\end{abstract}

\pacs{73.20.At,74.20.-z,72.80.Rj}
\maketitle

\draft
\preprint{Dec 07, 2001}

Depending on several parameters 
such as hopping integrals or chemical potentials, 
and also on underlying crystalline lattices,
a large variety of electronic structures are
realized in condensed matter physics.
Electron correlations
also give rise to a plenty of quantum phases,
forming non-trivial quasi-particle band structures.
An interesting consequence of 
a rich band structure
is the existence of edge states
that may appear 
when boundaries are present.
In the quantum Hall effect (QHE), this issue was discussed 
in terms of the origin of the quantization of 
a Hall conductance.
\cite{laughlin,halperin,thouless,kohmoto,hatsugai}
Recently, the ideas developed in the QHE
have also been extended 
for other gapped many-body systems,
and become essential 
to describing topological nature 
of several quantum phases. 
\cite{volovik,senthil,morita,read,goryo,hatsugai2}

Apart from these examples for gapped systems,
edge states in gapless systems have 
attracted much attention recently.
Example of these are
$d$-wave superconductor (SC) with edges \cite{hu,adagideli},
or
graphite ribbons\cite{fujita},
where
the existence of edge states
strongly depends on the shape of edges.
For $d$-wave SC with edges,
the zero bias conductance peak (ZBCP)
due to zero-energy edge states
was observed via a tunneling spectroscopy.
\cite{tanaka,iguchi}

The issue addressed
in this Letter is how to infer
the existence of zero-energy eigen states
localized on the boundaries in terms of 
properties of the bulk,
and the symmetry.
We first consider one-dimensional (1D) systems
with a particle-hole symmetry,
and then apply
the results to systems
in higher dimensions.
Especially, we will demonstrate applications to 
fully gapped SC in conjunction with
the Chern number,
2D $d$-wave SC,
and graphite ribbons.
In addition to these examples,
the present work is also applicable
to zero-modes in the 1D molecule polyacetylene \cite{su},
and quantum spin systems.

We start with 
the following single-particle Hamiltonian on a 1D lattice:
\begin{eqnarray}
{\mathcal H}=
\sum_{x,x'}
\boldsymbol{c}_{x}^{\dagger}h_{x,x'}\boldsymbol{c}_{x'}^{},
&&
h_{x,x'}=
\left[
\begin{array}{cc}
t_{x,x'} & \Delta_{x,x'} \\
\Delta^{\prime}_{x,x'} & -t_{x,x'}
\end{array}
\right]=h_{x',x}^{\dagger},
\nonumber 
\end{eqnarray}
where 
$
t_{x,x'},\Delta_{x,x'},\Delta'_{x,x'}\in \mathbb{C}
$,
and
$\boldsymbol{c}^{\dagger}_{x}=(c^{\dagger}_{x\uparrow},c_{x\downarrow})$
denotes electron creation/annihilation operators
at site $x$.
The total number of the lattice sites is $N_{x}$
and $x=1,\cdots , N_{x}$.
This Hamiltonian includes
the Bogoliubov-de Gennes (BdG) Hamiltonian
both for singlet and ( some of ) triplet SC.

In the following, we consider two types of Hamiltonians: 
the bulk Hamiltonian and the edge Hamiltonian.
As for the bulk Hamiltonian,
assuming that the system is translationally invariant,
$h_{x,x'}=h(x-x')$,
and adopting periodic boundary condition (PBC),
we can perform the Fourier transformation to obtain
$
{\mathcal H}^{\mathrm{bulk}}=
\sum_{k}
\boldsymbol{c}_{k}^{\dagger}h_{k}\boldsymbol{c}_{k}^{}
=
\sum_{k}\boldsymbol{c}_{k}^{\dagger}
\left[
\begin{array}{cc}
\xi_{k}        & \Delta_{k} \\
\Delta_{k}^{*} & -\xi_{k}
\end{array}
\right]
\boldsymbol{c}_{k}^{}
$,
where
$
\boldsymbol{c}_{x}=1/\sqrt{N_{x}}\sum_{k}e^{ikx}\boldsymbol{c}_{k}
$,
$k\in (-\pi,\pi]=S^{1}$ is the crystal momentum,
and 
$\xi_{k}\in \mathbb{R},\Delta_{k}\in \mathbb{C}$.
Since $\left( \sigma_{Y} h_{k} \sigma_{Y} \right)^{*}=-h_{k}$
($\sigma_{X,Y,Z}$ represent the Pauli matrices),
eigenvalues $E$ and $-E$ always appear in pair
for each $k$, 
which we call the particle-hole symmetry.
Let us introduce
a convenient parametrization for ${\mathcal H}^{\mathrm{bulk}}$
in $k$-space
\begin{eqnarray}
h_{k}=\boldsymbol{R}(k)\cdot \boldsymbol{\sigma},
\nonumber 
\end{eqnarray}
where $\boldsymbol{R}(k)
=(X,Y,Z)
:=(\mbox{Re}\Delta_{k},-\mbox{Im}\Delta_{k},\xi_{k})
\in \mathbb{R}^{3}$.
In this parameterization, 
the energy eigenvalues are given by
$E(k)=\pm |\boldsymbol{R}(k)|$.
The origin ${\mathcal O}\in \mathbb{R}^{3}$ 
corresponds to the gap-closing point.
For a given $k\in S^{1}$,
there exists a one to one correspondence between
a point in a 3D space
$\boldsymbol{R}(k)$ and $h_{k}$,
and hence
we can identify a loop $\ell:
k\in S^{1}\rightarrow \boldsymbol{R}(k)\in \mathbb{R}^{3}$
for each 1D Hamiltonian ${\mathcal H}^{\mathrm{bulk}}$;
for a given parametrized loop,
we can always reconstruct ${\mathcal H}^{\mathrm{bulk}}$
by inverse Fourier transformation.
\cite{rem0}
We write
${\mathcal H}^{\mathrm{bulk}}[\ell]$
for the Hamiltonian which corresponds
to $\ell$ hereafter.

An edge Hamiltonian ${\mathcal H}^{\mathrm{edge}}$ is generated
by truncating a bulk Hamiltonian ${\mathcal H}^{\mathrm{bulk}}[\ell]$
in a certain way.
We refer an edge Hamiltonian as
${\mathcal H}^{\mathrm{edge}}[\ell,e]$,
where $e$ represents a prescription for creating edges.
For example,
a natural way of truncation is to 
prohibit all the matrix elements across $N_{x}$,
i.e., set $h_{x,x'}=0$ if $N_{x}\in [x,x']$,
which we call $e_{c}$.
Generally, $e$ can represent an impurity potential at an edge, 
coexistence of different order parameters near boundaries
in superconducting systems,
etc.
Then, we ask if
${\mathcal H}^{\mathrm{edge}}[\ell,e]$ supports
zero-energy states localized at either end of the sample
for given $\ell$ and $e$.
Our strategy to answer this question 
is to consider a 
continuous deformation of a Hamiltonian 
from a reference Hamiltonian with
exact zero-energy edge states,
in conjunction with a symmetry.

In the following,
let us focus on a loop on a 2D plane that contains the origin ${\mathcal O}$
in $\boldsymbol{R}$-space.
We crown such loops 
with a superscript $^{*}$ as a reminder,
thereby referred to as $\ell^{*}$.
As a prescription for creating edges, we adopt $e_{c}$
for a while.
Let $|\left.\ell,E,p\right>$
denote an edge states of ${\mathcal H}^{\mathrm{edge}}[\ell,e_{c}]$
with energy $E$,
localizing at $p=L(R)$ 
where $L(R)$ represents the left(right) edge.
We assume a state which appears
within the bulk energy gap is
localized at either end of the sample
for an infinite system.
A state localized at both ends also may appears,
which is a superposition made from
two independent edge states localized at 
the left and right.
We will show
\begin{quote}
(A)
if 
${\mathcal H}^{\mathrm{edge}}[\ell^{*},e_{c}]$
has an edge state at non-zero energy
$|\left.\ell^{*},E\neq 0,p\right>$,
it also has
$|\left.\ell^{*},-E,p\right>$
which localizes at the same edge,
with the opposite energy.
\end{quote}

First, note that
we can restrict ourselves to loops on the $XY$-plane,
since an arbitrary 2D plane can be rotated
to the $XY$-plane by a unitary transformation:
a global $SO(3)$ rotation in $\boldsymbol{R}$-space,
which amounts to a $SU(2)$ transformation 
on $\boldsymbol{c}_{x}$
for {\it each site}.
To prove the statement,
it is essential that
the particle-hole symmetry is promoted to 
the chiral symmetry 
for ${\mathcal H}^{\mathrm{edge}}[\ell^{*},e_{c}]$.
Since all the hopping $t_{x,'x}$ is zero for loops on the $XY$-plane,
the Hamiltonian can be expressed as
${\mathcal H}^{\mathrm{edge}}[\ell^{*},e_{c}]=
(\boldsymbol{c}^{\dagger}_{\uparrow},\boldsymbol{c}_{\downarrow}^{})H
\left(
\begin{array}{c}
\boldsymbol{c}_{\uparrow}^{}\\
\boldsymbol{c}_{\downarrow}^{\dagger}
\end{array}
\right)
$,
$
H=\left[
\begin{array}{cc}
{\bf 0} & {\mathcal D}^{} \\
{\mathcal D}^{\dagger} & {\bf 0}
\end{array}
\right]
$.
Then,
$\Gamma:={\bf 1} \otimes \sigma_{Z}$ anticommutes with $H$,
$\Gamma H \Gamma =-H$,
which we call the chiral symmetry.
Consequently,
if ${\mathcal H}^{\mathrm{edge}}[\ell^{*},e_{c}]$ 
has an edge mode 
$|\left. \psi \right>=|\left.\ell^{*},E\neq 0,p\right>$,
it also has an edge mode 
with energy $-E$, 
$\Gamma|\left. \psi \right>=|\left.\ell^{*},-E,q\right>$.
Moreover, since $\Gamma$ is a purely local operator
which only changes the phase of $c_{\uparrow}$,
i.e.,
it does not ``mix''
the coordinate in the real space,
$|\left. \psi \right>$
and
$\Gamma|\left. \psi \right>$
should be localized at the same edge,
$q=p$.
Notice the above discussion is not applicable
for $E=0$,
since both $|\left. \psi \right>$
and $\Gamma|\left. \psi \right>$
have the same energy,
and hence can be the equivalent state.

\begin{figure}
\begin{center}
\includegraphics[width=8.5cm,clip]{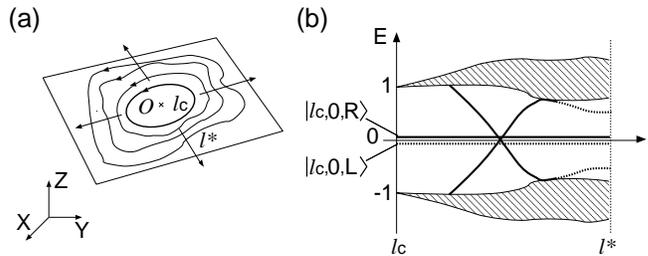}
\caption{
\label{deform}
(a) Continuously deforming 
$\ell_{c}$ into a loop $\ell^{*}\sim \ell_{c}$.
During the deformation,
the loop is kept on the 2D plane,
without crossing ${\mathcal O}$.
(b)
A possible energy spectrum during the deformation.
A thick/broken line represents 
a edge mode localized at $R/L$.
}
\end{center}
\end{figure}

Next, we further assume that
$\ell^{*}$ is continuously
deformed into a unite circle $\ell_{c}$
centered at ${\mathcal O}$,
such that
the loop is always on the 2D plane,
and 
does not cross ${\mathcal O}$
during the deformation.
(Fig. \ref{deform}).
For a loop $\ell^{*}$ with this property,
we write as $\ell^{*}\sim \ell_{c}$ henceforth.
We can prove that:
\begin{quote}
(B)
${\mathcal H}^{\mathrm{edge}}[\ell^{*}\sim \ell_{c},e_{c}]$ has at least
a pair of edge states at zero energy.
\end{quote}

To see this, 
we focus on
$\boldsymbol{R}(k)=(
\cos k,-\sin k,0 )
$
and the corresponding Hamiltonian 
$
{\mathcal H}^{\mathrm{edge}}[\ell_{c},e_{c}]
=\sum_{x=1}^{N_{x}-1}
\boldsymbol{c}^{\dagger}_{x}
\left[
\begin{array}{cc}
0 & 1 \\
0 & 0
\end{array}
\right]
\boldsymbol{c}_{x+1}^{}
+
h.c.
$
Since $c_{1\downarrow}^{},c_{1\downarrow}^{\dagger},
c_{N_{x}\uparrow}^{},c_{N_{x}\uparrow}^{\dagger}$
do not appear in $
{\mathcal H}^{\mathrm{edge}}[\ell_{c},e_{c}]$,
there are two exact zero-energy levels,
which localize at $x=1$ and $x=N_{x}$,
i.e.,
${\mathcal H}^{\mathrm{edge}}[\ell_{c},e_{c}]$ has
two edge states
$
|\left. \ell_{c},0,L \right>
$
and 
$
|\left. \ell_{c},0,R \right>
$.
\cite{rem1}
By assumption,
we can deform $\ell_{c}$ into $\ell^{*}$
continuously.
During the deformation,
$
|\left. \ell_{c},0,L \right>
$
and 
$
|\left. \ell_{c},0,R \right>
$
do not go away from zero energy,
since we can apply (A),
and the bulk energy gap does not collapse.
Although other edge states
$
|\left. E,p \right>
$
and 
$
|\left.-E,p \right>
$
may appear in pair from the bulk energy bands,
since the number of edge modes
localized at $L/R$ is always odd,
there must exist at least a pair of zero-energy states.

Although we have concerned ourselves with
a certain type of edges $e_{c}$, 
let us next consider to adiabatically modify $e_{c}$.
As far as the modified prescription 
does not break the chiral symmetry,
the perturbed Hamiltonian also supports exact zero-energy edge states,
since perturbations at the edges do not collapsed the energy gap.
Thus, we have showed
\begin{quote}
(C)
for a prescription $e^{*}$ that respects the chiral symmetry,
${\mathcal H}^{\mathrm{edge}}[\ell^{*}\sim \ell_{c},e^{*}]$
possesses at least a pair of zero-energy states.
\end{quote}

In summary,
there are three conditions 
for ${\mathcal H}^{\mathrm{edge}}[\ell,e]$
to support zero-energy edge states:
(A) $\ell$ is on a 2D plane that contains ${\mathcal O}$,
($\ell^{*}$),
(B) $\ell$ is continuously deformed to $\ell_{c}$
without crossing ${\mathcal O}$,
($\ell\sim \ell_{c}$),
and 
(C) $e$ respects the chiral symmetry,
($e^{*}$).

We have established our main results,
and a few comments are in order.
First, notice that 
the edge states discussed here are not at
exact zero energy for a finite system,
though 
$|\left. \ell_{c},0,L \right>$
and 
$|\left. \ell_{c},0,R \right>$
are exact zero-energy state.
This is allowed since
an assumption for the statement (A) does not hold
for a finite system size.
In this case,
a state localized at both ends
cannot be decomposed into two-independent edge states,
which we can regard as a hybridized state made from
the two edge modes at the left and right.
In $N_{x}\rightarrow \infty$,
this state becomes degenerate with 
another hybridized state.

Second,
consider a unit circle $\ell_{c}^{n}$ that
encloses ${\mathcal O}$ $n$ times ($n$:odd).
${\mathcal H}^{\mathrm{edge}}[\ell_{c}^{n},e_{c}]$ 
can be diagonalized
in the same way as ${\mathcal H}^{\mathrm{edge}}[\ell_{c},e_{c}]$ , resulting in
$2n$ exact zero-energy states.
Then, by the same discussion,
a class of Hamiltonians
${\mathcal H}^{\mathrm{edge}}[\ell^{*}\sim \ell_{c}^{n},e^{*}]$ have
at least one pair of edge states at $E=0$.

Finally,
the present discussion is 
consistent with
the King-Smith-Vanderbilt (KSV) formula
that relates 
macroscopic polarization
to
the Zak's geometric phase $\gamma$.
\cite{zak,kingsmith,resta,rem2}
However, the KSV formula does not tell us
if edge states are at zero energy.

We go on to applications of the present results.
We adopt $e_{c}$ as a prescription for creating edges 
unless otherwise stated.
First, we discuss fully gapped systems and
edge states.
Especially, we comment on
a topological aspect of 2D SC with a full gap,
whose examples include
$d+id$ SC and 
the chiral $p$-wave SC.
\cite{volovik,senthil,morita,read,goryo,hatsugai2}
For these SC, we can
define an integer called the Chern number,
non-zero value of which 
implies the existence of edge states
connecting the upper and the lower bands
as known in the QHE \cite{hatsugai}.
The present results are
consistent with this discussion.
For 2D systems with edges, 
we first Fourier transform 
along a direction parallel to the edge,
to get a family of 1D Hamiltonians
parametrized by the wave number along the edge.
Then, we can apply the present discussions
for each 1D Hamiltonian.
Since the non-zero Chern number implies
there exists a loop which 
is on a plane and encloses ${\mathcal O}$ \cite{hatsugai2},
both the topological argument and the present results 
lead to existence of zero-energy edge modes.
For fully gapped systems,
edge modes
are expected to be stable
even in the presence of electron-electron interaction
as far as the bulk energy gap is not collapsed.

\begin{figure}
\begin{center}
\includegraphics[width=8.5cm,clip]{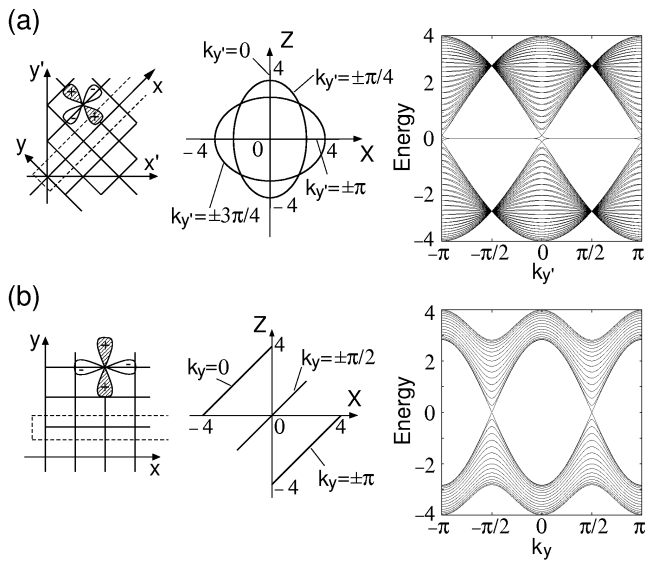}
\caption{
\label{d}
Loops in $\boldsymbol{R}$-space and 
the energy spectrum of $d_{x^{2}-y^{2}}$-wave SC with 
(a) $(110)$ and (b)$(100)$ surfaces.
Dotted squares show a choice of unit cell
in Fourier transforming along the edges.
The calculation is for $N_{x}=50$ for $(110)$ surfaces,
and $N_{x}=30$ for $(100)$ surfaces.
}
\end{center}
\end{figure}

\begin{figure}
\begin{center}
\includegraphics[width=8.5cm,clip]{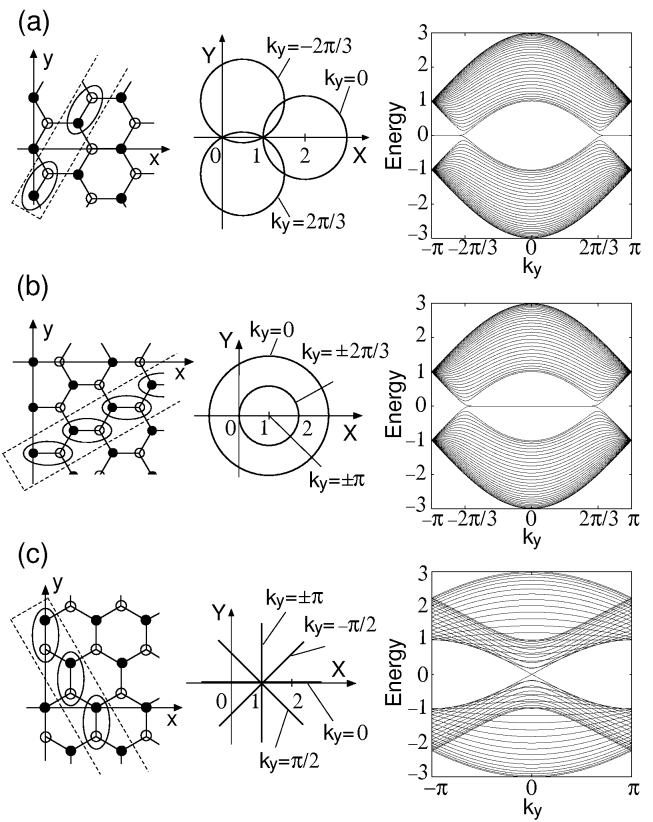}
\caption{
\label{graphite}
Loops in $\boldsymbol{R}$-space and
the energy spectrum of a graphite ribbon with 
(a) zigzag, (b) bearded, and (C) armchair edges.
The ovals 
indicate how to form a spinor $\boldsymbol{c}$,
for each edge,
and dotted squares show a choice of unit cell
in Fourier transforming along the edges.
The loops corresponding to a one-parameter family
of Hamiltonians
are (a)
$\boldsymbol{R}_{k_{y}}(k_{x'})=
(\cos (k_{y}-k_{x'})+1+\cos k_{y},
-\sin (k_{y}-k_{x'})+\sin k_{y},0)$,
(b)
$
(\cos k_{x'}+\cos (k_{y}-k_{x'})+1,
\sin k_{x'}-\sin (k_{y}-k_{x'}),0)$,
(c)
$(\cos (k_{x'}+k_{y})+\cos k_{x'}+1,
-\sin (k_{x'}+k_{y})+\sin k_{x'},0) $.
Here we have taken all the hopping integral equal to unity,
and $k_{y}$ is a wave number along the edges.
The calculation is for $N_{x'}=30$ for zigzag and bearded edges,
and $N_{x'}=29$ for armchair edges. 
}
\end{center}
\end{figure}

Although the topological argument 
is only applicable 
for fully gapped systems,
our results here are not restricted to gapped cases,
and can be applicable also for gapless cases in
arbitrary dimensions.
Here, as an application,
we consider surface states for $d_{x^{2}-y^{2}}$-wave SC.
In Ref. \cite{hu},
a semi-classical approach was employed to show
the sign change of the pair potential at a (110) surface
gives rise to existence of edge states,
which can be used as a phase sensitive probe
to detect pairing symmetries.
It was also pointed out the Andreev equation
for the present system is closely related
to Witten's supersymmetric quantum mechanics.
\cite{kosztin,adagideli}
Here, we discuss this issue
with a lattice regularization.

Consider 2D $d_{x^{2}-y^{2}}$-wave SC 
$
{\mathcal H}^{\mathrm{bulk}}=
\sum_{r}^{PBC}
\left[
\boldsymbol{c}^{\dagger}_{r}
h_{x}
\boldsymbol{c}_{r+x}^{}
+\boldsymbol{c}^{\dagger}_{r}
h_{y}
\boldsymbol{c}_{r+y}^{}
+h.c.
+\boldsymbol{c}^{\dagger}_{r}
h_{0}
\boldsymbol{c}_{r}^{}
\right],
$
where
$
h_{x}=
\left[
\begin{array}{cc}
t     & \Delta \\
\Delta & -t
\end{array}
\right]
$,
$
h_{y}=
\left[
\begin{array}{cc}
t     & -\Delta \\\
-\Delta & -t
\end{array}
\right]
$,
and
$
h_{0}=
\left[
\begin{array}{cc}
\mu  & 0\\
0    & -\mu
\end{array}
\right]
$.
(We set $t=\Delta=1,\mu=0$ as an example.)
We terminate this system,
and consider $(110)$ surfaces first.
Fourier transforming along the $y'$ direction 
in Fig. \ref{d} (a),
we obtain a family of 1D Hamiltonians
parametrized by $k_{y'}$.
The corresponding loops are
$
\boldsymbol{R}_{k_{y'}}(k_{x})=
\left(
2\cos (k_{x}-k_{y'})-2\cos k_{x},
0,
2\cos (k_{x}-k_{y'})+2\cos k_{x}
\right)$.
For a given $k_{y'}$,
$
(1+\cos k_{y'})(X/2)^{2}+
(1-\cos k_{y'})(Z/2)^{2}
=2\sin^{2} k_{y'}
$
is satisfied,
which is an ellipsis on the $XZ$-plane enclosing ${\mathcal O}$.
Thus, from the above discussion,
the present system supports
zero-energy surface states
for all $k_{y'}$ 
except at the gap-closing points
$k_{y'}=\pm\pi,0$
where the loop collapses into a line segment.

On the other hand,
for (100) surfaces,
we obtain
$
\boldsymbol{R}_{k_{y}}(k_{x})
=
(
2( \cos k_{x}-\cos k_{y} ),
0,
2(\cos k_{x}+\cos k_{y})
)$,
which is a line segment on the $XZ$-plane
for all $k_{y}$.
Zero-energy edge states are not expected to exist
for this case.
We have verified numerically this 
prediction in 
Fig. \ref{d}(b).

Let us comment on an interplay between
zero-energy edge states and interactions
for the present case.
If we treat the problem self-consistently,
coexistence of $is$- or $id_{xy}$-wave order parameter
with $d_{x^{2}-y^{2}}$-wave 
near the surface is possible for the $(110)$ surface,
locally breaking the time-reversal symmetry.\cite{matsumoto}.
This can be interpreted based on 
the present discussions as follows.
Since edge states with different $k_{y^{\prime}}$ are 
all degenerate at $E=0$,
they are expected to cause a Peierls-like instability.
In presence of interactions,
parameters in a single particle Hamiltonian
$t,\Delta,\Delta^{\prime}$ near the edges
might be effectively modified
in order to lift the degeneracy,
and thereby lower the ground state energy.
However, since these zero-energy edge states are stable 
to perturbations which respect the chiral symmetry
(statement (C)),
such modifications should be accompanied 
with the breaking of the chiral symmetry
near the boundaries.
The emergence of $is$ or $id_{xy}$ components 
near the boundary indeed breaks the chiral symmetry
to lift the degeneracy of edge modes,
while a purely real order parameter cannot do it.

We turn to edge states in graphite ribbons.
There are several types of edges
for a graphite ribbon such as
zigzag, bearded, and armchair edge.\cite{fujita}
Defining
$c_{\uparrow}^{}=c_{\bullet}$ and 
$c_{\downarrow}^{\dagger}=c_{\circ}$,
where $c_{\bullet / \circ}$ is
an electron annihilation operator
on a sublattice $\bullet / \circ$,
we can apply our formalism
to graphite ribbons.
Notice that
we have several options
for choosing 
$c_{\bullet / \circ}$
to form a spinor
$\boldsymbol{c}$,
since they live on {\it different} sites.
When we truncate the system,
these choices 
lead to different shapes of edges (Fig. \ref{graphite}).
Taking an appropriate pair
for each type of edge as indicated in Fig. \ref{graphite},
we can discuss in parallel
to the above SC example.
The existence of zero-energy edge states 
is predicted 
for the zigzag and the bearded case,
while we do not expect zero-energy edge states for an armchair edge,
which is confirmed by a numerical calculation
(see Fig.\ref{graphite}).
These zero-energy edge modes are continuously 
connected to the gapless bulk spectrum,
forming a flat band and a sharp peak in density of states at the fermi energy.
This might trigger an instability
in presence of electron-electron or electron-phonon interactions,
which leads to, for example,
a magnetic polarization 
near the boundaries.
\cite{fujita}

To conclude,
we have established a criterion to determine 
the existence of zero-energy edge modes
in terms of bulk properties and the chiral symmetry.
Our strategy is 
to make use of the chiral symmetry,
and a 
continuous deformation 
of a reference Hamiltonian with
exact zero-energy edge states.
The present discussions are applicable for 
both gapped and gapless systems
in arbitrary dimensions.

We thank Y. Morita, C. Mudry, and K. Kusakabe
for fruitful discussions.
S.R. is grateful to T. Oka and K. Nomura
for useful comments.


\begin{references}
\bibitem{rem0}
Note that that for a finite system with PBC,
the loop is actually a discrete set of points.
\bibitem{rem1}
${\mathcal H}^{\mathrm{edge}}[\ell_{c},e_{c}]$ can be diagonalized as
$
{\mathcal H}^{\mathrm{edge}}[\ell_{c},e_{c}]
=\sum_{x=1}^{N_{x}-1}
[
d^{\dagger}_{+,x+1/2}d_{+,x+1/2}^{}-d^{\dagger}_{-,x+1/2}d_{-,x+1/2}^{}
]
$
by a unitary transformation
$
d_{\pm,x+1/2}^{\dagger}:=
(
c_{x\uparrow}^{\dagger}\pm c_{x+1\downarrow}^{}
)/\sqrt{2}
$.
\bibitem{rem2}
For all loops on a 2D plane that contains ${\mathcal O}$,
$\gamma$ is equal to an integral multiple of $\pi$,
since it corresponds to half of the solid angle
subtended by the loops.\cite{berry}
\end{references}
\end{document}